\documentclass[pss]{wiley2sp} 
\usepackage{amsmath}
\usepackage{braket}
\usepackage{textcomp}

\newcommand{\siv}{SiV\textsuperscript{-} }

\begin{document}

\title{Coherence properties and quantum control of silicon vacancy color centers in diamond}

\titlerunning{Coherence properties and quantum control of silicon vacancy color centers in diamond}

\author{%
  Jonas Nils Becker\textsuperscript{\textsf{\bfseries 1,2}},
  Christoph Becher\textsuperscript{\Ast,\textsf{\bfseries 1}},
}

\authorrunning{J. N. Becker et al.}

\mail{e-mail
  \textsf{christoph.becher@physik.uni-saarland.de}, Phone:
  +49-681-3022466, Fax: +49-681-3024676}

\institute{%
  \textsuperscript{1}\,Universit\"at des Saarlandes, Naturwissenschaftlich-Technische Fakult\"at, Campus E2.6, 66123 Saarbr\"ucken, Germany\\
    \textsuperscript{2}\,Clarendon Laboratory, University of Oxford, Parks Road, Oxford OX1 3PU, United Kingdom\\
}



\abstract{%
%
%
%
\abstcol{
Atomic-scale impurity spins, also called color centers, in an otherwise spin-free diamond host lattice have proven to be versatile tools for applications in solid-state-based quantum technologies ranging from quantum information processing (QIP) to quantum-enhanced sensing and metrology. Due to its wide band gap, diamond can host hundreds of different color centers. However, their suitability for QIP or sensing applications has only been tested for a handful of these, with the nitrogen vacancy (NV) strongly dominating this field of research. Due to its limited optical properties, the success of the NV for QIP applications however strongly depends on the devel-
  }{
  opment of efficient photonic interfaces. In the past years the negatively charged silicon vacancy (SiV) center received significant attention due to its highly favourable spectral properties such as narrow zero phonon line transitions and weak phonon sidebands. We here review recent work investigating the SiV centre's orbital and electron spin coherence properties as well as techniques to coherently control its quantum state using microwave as well as optical fields and we outline potential future experimental directions to improve the SiV's coherence time scale and to develop it into a valuable tool for QIP applications.
 	}}

%
%

\maketitle   

\section{Introduction}

\begin{figure}[t]%
	\includegraphics*[width=\linewidth]{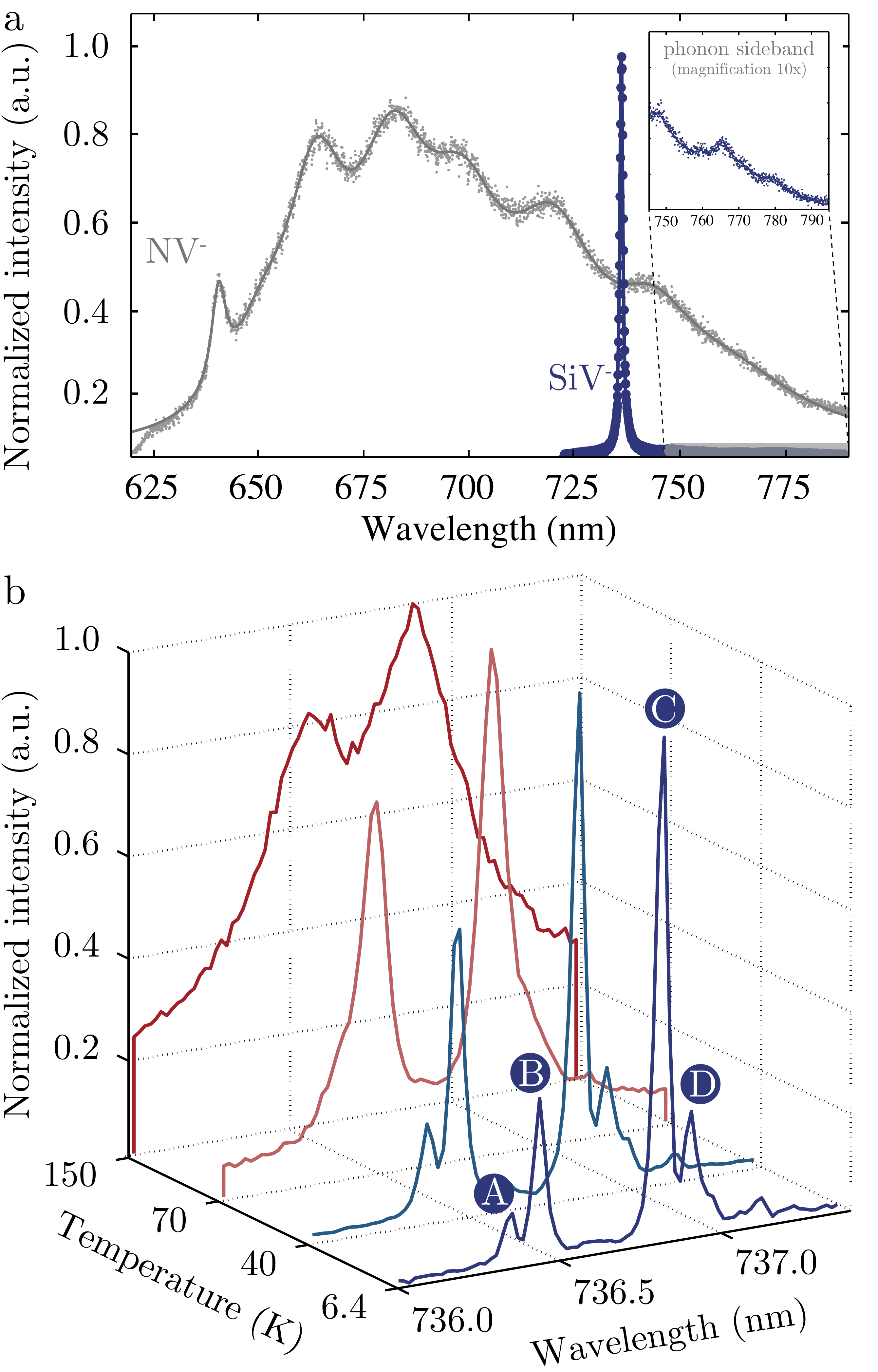}
	\caption{(a) Room temperature fluorescence spectrum of the \siv zero phonon line (blue lines and dots) and its weak phonon sideband (insert). The broad emission of the NV\textsuperscript{-} is shown in the background (grey) for comparison \cite{beckerdiss}. (b) Low-temperature spectra of the \siv zero phonon line for several temperatures between 150\,K and 5\,K showing the emergence of the characteristic four-line finestructure \cite{beckerdiss}.}
	\label{fig1}
\end{figure}

Color centers in diamond, seen as a general class of confined impurity spins in a spin-free host material, offer a multitude of opportunities in spin physics ranging from fundamental studies of engineered mesoscopic spin system dynamics to potential applications emerging from quantum control  \cite{awschalom2013}. In contrast to their counterparts in silicon, diamond-based impurities can be optically active in the visible to near-infrared spectral region, efficiently detectable with modern single-photon counting systems \cite{prawer}. The most prominent example for such a system is the nitrogen vacancy (NV) center \cite{doherty2013,acosta2013}, which has been studied extensively throughout the past decade. Due to its favourable electronic structure featuring a spin-split ground state accessible by microwave fields, very long spin coherence times and an internal dynamics that conveniently allows for spin initialization and read-out using off-resonant lasers, a large number of pioneering demonstrations in quantum information applications have been carried out using this center \cite{childress2013}. Beyond microwave control there have been a few demonstrations of all-optical spin manipulation of the NV center \cite{santori2006,bassett2014}, which however requires the application of additional perturbations such as strain or electric fields to mix the excited state components and enable optical access to spin-flipping transitions. The most crucial drawback of the NV however is the very limited quality of its optical spectrum, which is dominated by a very broad phonon sideband, hindering efficient optical spin access and interfacing of the spin with photons. The low emission rate into its zero phonon line (ZPL) e.g severely restricts the attainable rates in entanglement generation schemes \cite{hensen2015} and hence limits the scalability of quantum processors and networks based on NV centers. Therefore, the current challenges in the field of diamond-based quantum technologies are either to amplify the zero-phonon emission of the NV, e.g. by coupling it to optical cavities \cite{faraon2012,li2015,aharonovich2014,schroder2016} or to check the wide range of hundreds of colour centres for novel spin impurities which combine good photonic qualities with decent spin properties.\\
One particularly promising alternative colour centre is the negatively charged silicon vacancy center (SiV\textsuperscript{-}) \cite{wang2006}. Over the past couple of years it received steadily increasing attention due to its excellent spectral properties \cite{sipahigil2014,neu2011}. This includes room-temperature ZPL linewidths of less than 1\,nm \cite{neu2011,neu20112} as displayed in Fig.\,\ref{fig1}(a) and almost lifetime-limited linewidths at liquid helium temperatures \cite{rogers20143} without signs of spectral diffusion, small phonon sidebands with Huang-Rhys factors of down to S=0.08 \cite{neu2011} as well as narrow inhomogeneous distributions of about 10\,GHz in dense ensembles \cite{sternschulte1994} down to 300\,MHz in less dense samples \cite{sipahigil2014}. These features for example allow the generation of indistinguishable photons from separate \siv centres without the need for additional tuning techniques and hence enable applications as an excellent single photon source or as a building block for scalable quantum networks \cite{sipahigil2014}. However, for applications in quantum information processing and quantum communication the center, for a long time, lacked techniques to coherently manipulate its quantum states. This coherent control has only very recently been added to the \siv toolbox and this article provides a comprehensive review of these experiments and a compact introduction to the most relevant physics of the \siv center which, in many aspects, differs significantly from the one of its well-studied counterpart, the NV center.
\section{Electronic properties}
\subsection{Level structure}
\begin{figure*}[htb]%
	\includegraphics*[width=\textwidth]{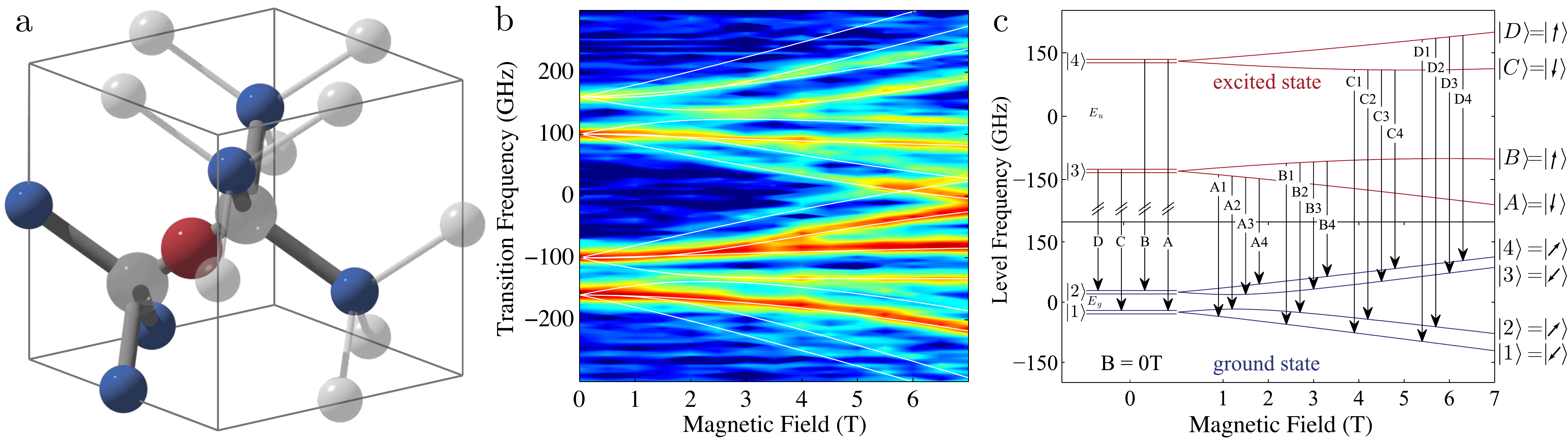}
	\caption{(a) Molecular structure of the \siv featuring a silicon atom (red), two vacancies (dark grey) and six nearest-neighbour carbon atoms (blue) embedded in the diamond structure (light grey) \cite{beckerdiss}. (b) Fluorescence spectra of the \siv ZPL for variable magnetic field strengths. White lines correspond to simulations with a group-theoretical model \cite{hepp2014}. (c) Electronic level structure of the \siv at zero and finite magnetic field \cite{hepp2014}.
	}
	\label{fig2}
\end{figure*}

Before focusing on the quantum control, we first review the \siv center's electronic structure as it is key to understand the center's coherence properties as well as the employed control techniques. Ab initio calculations of the center's molecular structure predict a trigonal-antiprismatic (point group D$_\text{3d}$) as displayed in Fig.\,\ref{fig2}(a) with the silicon atom, presumably due to its size, relaxing onto an interstitial lattice site in between two empty carbon sites along $\langle111\rangle$ and surrounded by six equivalent nearest-neighbour carbon atoms \cite{goss1996}. This orientation is strongly supported by polarization measurements in unstrained bulk diamonds \cite{hepp2014,rogers20142}. Using this geometry as a starting point, group theory predicts a twofold spin and twofold orbitally degenerate $E_g$ ground  and $E_u$ excited state with only a single optical transition linking them \cite{hepp2014}.
To reproduce the four-line fine structure of the \siv displayed in Fig.\,\ref{fig1}(b) additional interactions due to spin orbit coupling and the Jahn-Teller effect have to be included. These perturbations lift the orbital degeneracy leading to an orbitally split ground and excited state with the characteristically large splittings of $\delta_g=2\pi\cdot 48\,GHz$ in the ground and $\delta_g=2\pi\cdot 259\,GHz$ in the excited state. By adding an additional Zeeman interaction term to the model and fitting it to fluorescence maps of single \siv centers acquired in external magnetic fields like the one depicted in Fig.\,\ref{fig2}(b), spin orbit coupling can be identified as the dominant interaction in this system. Moreover, the emergence of four optical components from each of the four zero-field transitions of the ZPL indicates an electronic spin of S=1/2 as each orbital ground and excited state component splits into two spin components. This is consistent with ESR investigations of the uncharged SiV$^0$ identifying it as a S=1 system \cite{edmonds2008,ulrika2011}. Note that, to comply with the existing literature, we here use a notation for the states that can be confusing. At zero magnetic field, the orbital ground states are labeled $\ket{1}$ and $\ket{2}$, the orbital excited states $\ket{3}$ and $\ket{4}$ and the four optical transitions A, B, C, D. At finite magnetic field we however label the ground states $\ket{1}...\ket{4}$, the excited states $\ket{A}...\ket{D}$ and we refer to the individual optical transitions via their initial and final states, e.g. A1 for the transition between $\ket{A}$ and $\ket{1}$. One circumstance that will be of importance for the optical control techniques discussed below is the fact that all 16 optical transitions, even those between spin sublevels of opposite spin projection are visible in the spectrum. This is due to a misalignment of the magnetic field aligned along (001) with respect to the \siv high symmetry axis by 54.7$^\circ$, leading to off-diagonal terms in the Zeeman Hamiltonians of ground and excited state, which mix individual spin-up and spin-down basis states. Since this mixing is different for ground and excited state it creates an overlap between levels of opposite spin projection, hence rendering the respective optical transitions weakly allowed. Alternatively, this can be understood as a difference in quantization axes of ground and excited states due to a competition of the external magnetic field, attempting a quantization along (001), and the spin orbit coupling forcing a quantization along (111). According to their relative strengths this results in an effective quantization axis slightly tilted away from (111) with the tilt angle being different in the ground and excited state due to different spin orbit strengths. These different quantization axes again cause an overlap of ground and excited state sublevels of opposite spin projections. The finite strengths of these transitions will allow coupling of ground state spin sublevels with opposite spin orientations to a common excited state, enabling optical spin control.
\subsection{Electron-phonon coupling}
In addition to the above discussed optical transitions, the \siv also features phonon-mediated transitions within its orbital doublet ground and excited state manifolds. This has first been observed for the excited states by resonantly exciting one of the excited state spin sublevels while measuring the resulting ZPL emission spectrum \cite{mueller2014}. Excitation into levels $\ket{C}$ or $\ket{D}$ of the upper excited state orbital branch results only in transitions originating from $\ket{C}$ and $\ket{A}$ or $\ket{D}$ and $\ket{B}$, respectively. This infers a spin state-preserving relaxation process between the excited state magnetic sublevels, faster than the excited state lifetime of about 1.8ns. Furthermore, the absence of transitions originating from $\ket{C}$ or $\ket{D}$ upon excitation of $\ket{A}$ or $\ket{B}$ indicates a thermally activated process as expected for phonon induced dynamics. Similar but slower processes have been identified within the ground state manifold by optical pumping experiments and using a microscopic model of the electron-phonon processes have been identified as first-order phonon transitions due to coupling of twofold-degenerate (transverse) acoustic phonons to the orbital doublet states \cite{jahnke2015}. At zero magnetic field the application of a laser pulse resonant with transition D leads to a transfer of population from state $\ket{2}$ into the lowest ground state during the pulse leading to a rising edge peak in the measured fluorescence response as depicted in the insert of Fig.\,\ref{fig3}(a). Due to a finite thermalization rate between both ground states, the population in state $\ket{2}$ then slowly recovers after the first pulse and this recovery can be monitored by the recovery of the rising edge peak of second pulse with delay $\tau$ yielding an orbital ground state relaxation time of T$_1^\text{orbit}$=39\,ns at 5\,K. A similar experiment can be repeated at finite magnetic field to determine the ground state spin relaxation \cite{rogers2014}. For this a first pulse applied to a "spin-flipping" transition like D1 is applied to initialize the spin in state $\ket{2}$. The population of this state is then read out by applying a second pulse on a strong cycling transition D2 and monitoring its rising edge fluorescence peak which, in this case, is decaying for increasing pulse delays $\tau$, yielding a spin relaxation time of T$_1^\text{spin}$=60\,ns for a magnetic field misaligned with the \siv axis by 70$^\circ$ and T$_1^\text{spin}$=2.4\,\textmu s for fields aligned with the \siv high symmetry axis, indicating great potential for the center's spin coherence \cite{rogers2014}.
\section{Electron spin properties}
\subsection{Spin coherence times \& limitations}
\begin{figure*}[htb]%
	\includegraphics*[width=\textwidth]{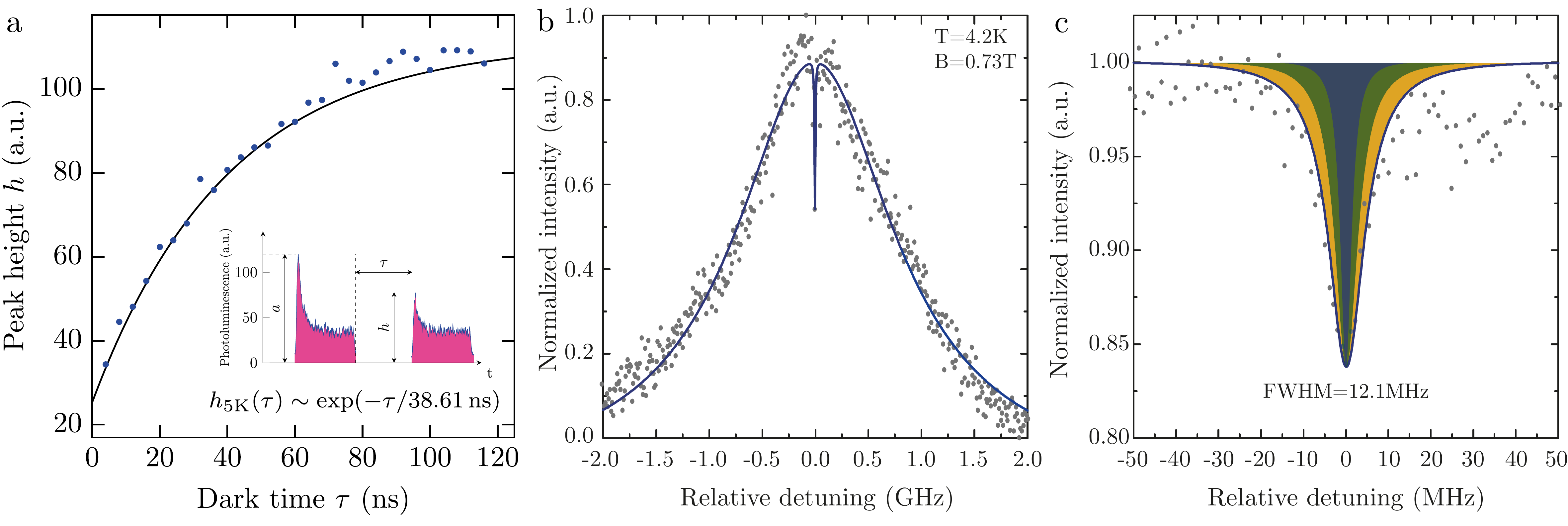}
	\caption{(a) Optical pumping measurement of the orbital coherence time T$_1^\text{orbit}$ of the \siv at zero magnetic field by applying two subsequent pulses resonant with transition D \cite{jahnke2015}. (b) Coherent population trapping at finite magnetic field by driving a $\Lambda$-scheme including transitions D1 and D2. In two-photon resonance a sharp dip corresponding to a coherent dark state is observed \cite{pingault2014}. (c) Zoom of the CPT dip and simulation using a four-level Bloch equation model. The model identifies a contribution of 3.5\,MHz of inherent spin decoherence (blue Lorentzian), 3.6\,MHz of power broadening (green) and 5.0\,MHz of limited relative laser coherence to the overall dip width of 12.1\,MHz \cite{pingault2014}.
	}
	\label{fig3}
\end{figure*}
The ground state spin coherence of the \siv has first been investigated using coherent population trapping (CPT) in external magnetic fields by applying two continuous wave lasers to two optical transitions linking ground state spin sublevels of opposite spin to a common excited state (a so-called $\Lambda$-scheme). If both beams are in two-photon resonance the system is pumped into a coherent dark state consisting of both ground state levels leading to a sharp dip in the fluorescence as depicted in Fig.\,\ref{fig3}(b). In this measurements the $\Lambda$-scheme is realized via transitions D1 and D2. The width of this dip is proportional to the lifetime of the superposition and hence the spin coherence time T$_2^*$. Using a density matrix model to include contributions from other decoherence sources such as power broadening and limited relative laser coherence, the authors of \cite{pingault2014} determined a contribution of the spin decoherence of 3.5\,MHz to the total dip width (blue Lorentzian in Fig.\,\ref{fig3}(c)) while power broadening and limited laser coherence contribute with 3.6\,MHz and 5.0\,MHz (green and yellow Lorentzians) to the total dip width of 12.1\,MHz. This decoherence rate corresponds to a coherence time of T$_2^*$=45\,ns whereas in a similar experiment \cite{rogers2014} a slightly lower coherence time of T$_2^*$=35\,ns has been measured, both orders of magnitude shorter than what has been reported for NV centers, even in nitrogen-rich diamond. Due to the similarity of this coherence time with the measured T$_1^\text{orbit}$ the authors of \cite{rogers2014} concluded that T$_2^*$ is limited by the above discussed phonon-driven transitions between orbital levels. Despite the fact that the random excitation from the spin sublevels of the lower into the higher orbital ground state branch is spin conserving, this process leads to an acquisition of random phase because of a slight difference in Zeeman splittings of both branches. However, while this process has only recently been verified to be the limiting factor at liquid helium temperatures for at least some \siv centers \cite{pingault2017}, in the original work the authors neglected the fact that the limit for the total coherence time in the absence of additional dephasing processes is given by T$_2^*$=2T$_1$. This indicates the presence of additional processes directly limiting the phase coherence (T$_2$) of the \siv such as electron or nuclear spin baths in the investigated samples (ion implanted type IIa HPHT diamond in \cite{pingault2014,pingault2017} and homoepitaxially grown Si-doped CVD diamond on type IIa HPHT substrate in \cite{jahnke2015,rogers2014}).
\subsection{Extending the spin coherence time}
\begin{figure}[t]%
	\includegraphics*[width=\linewidth]{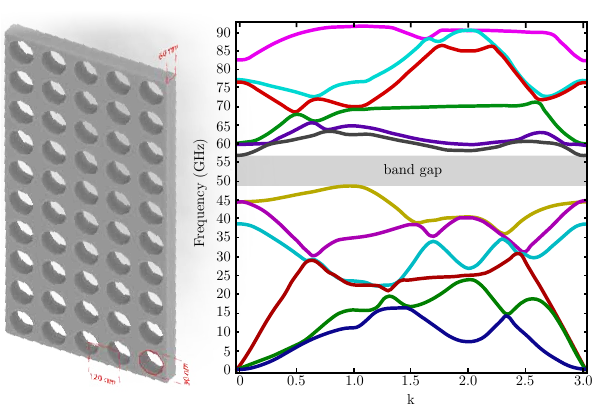}
	\caption{Simulated phononic band structure for a phononic band gap material displayed in the insert, featuring a 60\,nm thick diamond membrane structured with 90\,nm holes on a 120\,nm quadratic grid. A band gap around the \siv ground state splitting of 48\,GHz is clearly visible.}
	\label{fig4}
\end{figure}

Since the spin coherence of the SiV\textsuperscript{-}, a sample free of background spins provided, is limited by phonon-mediated processes, hope has been raised to significantly extent spin coherence by engineering the phonon environment of the center to suppress phonons with a frequency around 48\,GHz, corresponding to the center's orbital ground state splitting. This might for example be achieved by cooling the sample well below $\frac{48\text{GHz}}{k_B}$=2.3K, reducing the thermal population of the respective phonons and hence protecting the lowest ground state spin doublet from decoherence. An alternative approach is the fabrication of nanostructures small enough so that phonons of the respective frequency are suppressed \cite{albrecht2013b}. One way to achieve this is the fabrication of a phononic band gap by patterning diamond membranes to fabricate phononic crystals (in analogy to photonic crystals in which a band gap for photons is created) \cite{kipfstuhl2014}. However, first simulations, displayed in Fig.\,\ref{fig4}, show that, while these structures exist in theory, fabricating them is a challenging task using state-of-the-art diamond nanofabrication techniques since a full phononic band gap at 48\,GHz requires e.g. a quadratic grid (120nm spacing) of air holes with diameters of 90nm in only 60nm thick membranes. Hence, a more feasible approach might be the use of nanodiamonds with diameters smaller than 190\,nm, eliminating the respective phonons \cite{jahnke2015}. However, while the fabrication of such nanodiamonds is easily possible with current diamond growth techniques \cite{neu2011,neu20113} and although some \siv centers in such nanodiamonds show narrow low-temperature lines \cite{jantzen2016}, they currently provide strained crystal environments \cite{neu2011,neu20113} for the SiV\textsuperscript{-}, significantly influencing the center's electronic properties beyond acceptable margins for QIP applications. Alternatively, also the strain itself can be used to manipulate the phonon-mediated transition rates. Crystal strain is a purely orbital perturbation, increasing the ground and excited state splittings of the \siv and the corresponding phononic thermalization rate $\gamma_+$ and relaxation rate $\gamma_-$ according to \cite{jahnke2015} are given by
\begin{equation}
\gamma_+=2\pi\chi\rho\delta_{g}^3(exp(\frac{\hbar\delta_{g}}{k_BT})-1)^{-1}
\end{equation}
and
\begin{equation}
\gamma_-=2\pi\chi\rho\delta_{g}^3(exp(\frac{\hbar\delta_{g}}{k_BT}))^{-1}
\end{equation}
with the interaction frequency $|\overline{\chi_k(\omega)}|^2=\chi\omega$ of the phonon modes $k$ and the phonon density of states $\rho(\omega)=\rho\omega^2$. The resulting thermalization rate will therefore first increase cubically due to the increase in interaction frequency and phonon density of states but will for high enough splittings decay again due to the exponentially decreasing phonon occupation. This has first been observed in the distribution of optical linewidths for differently strained \siv centers implanted in type IIa HPHT diamond \cite{arend2016}. Recently, a more controlled approach for the application of crystal strain using a diamond nano-electro-mechanical system (NEMS) featuring single \siv centers in a metallically coated cantilever has been investigated\cite{sohn2017}. By applying a DC voltage, this system allows the application of a defined amount of strain to the \siv by bending the cantilever. With this method the authors were able to increase the \siv ground state splitting up to 470\,GHz, yielding an improved spin relaxation time of about T$_1^{spin}$=2.5\textmu s and a spin coherence time of T$_2^*$=250\,ns at 4\,K, limited by electrical breakdown of the NEMS device. Hence, reaching coherence times in the ms regime as they are e.g. needed for long distance quantum communication applications does not appear feasible using this technique and working at millikelvin temperatures, albeit still being technologically challenging, appears as the most feasible technique to potentially reach long coherence times with \siv centers.
\section{Coherent control}
A full set of single qubit operations as well as a universal two-qubit gate such as the controlled-NOT gate (CNOT) \cite{neumann2010b} between pairs of qubits is necessary to utilize the \siv in QIP applications. Coherent control of the spin and orbital degree of freedom of single \siv centers has recently been demonstrated \cite{becker2016,zhou2017,pingault2017}. In general, to achieve full coherent control, enabling preparation of arbitrary superposition states, rotations around two orthogonal axes of the Bloch sphere, i.e. the state space of a two-level qubit, are necessary. This is achieved by driving Rabi oscillations to control the relative amplitudes of both states in a superposition (also called angular control) and by Ramsey interference, utilizing the Larmor-precession of the superposition state during a defined free evolution time to set the relative phase of both state components (also called axis control).
\subsection{Microwave-based control}
The simplest method to control the state of the electron spin in the \siv is the application of a microwave resonantly driving the magnetic dipolar transition between two spin sublevels in the \siv ground state. By adjusting the strength of the magnetic field applied to lift the spin degeneracy the spin splitting and thus the microwave frequency can be conveniently tuned in the range of a few GHz, commonly used in other solid-state systems like the NV center to achieve similar types of control.
\begin{figure}[b]%
	\includegraphics*[width=\linewidth]{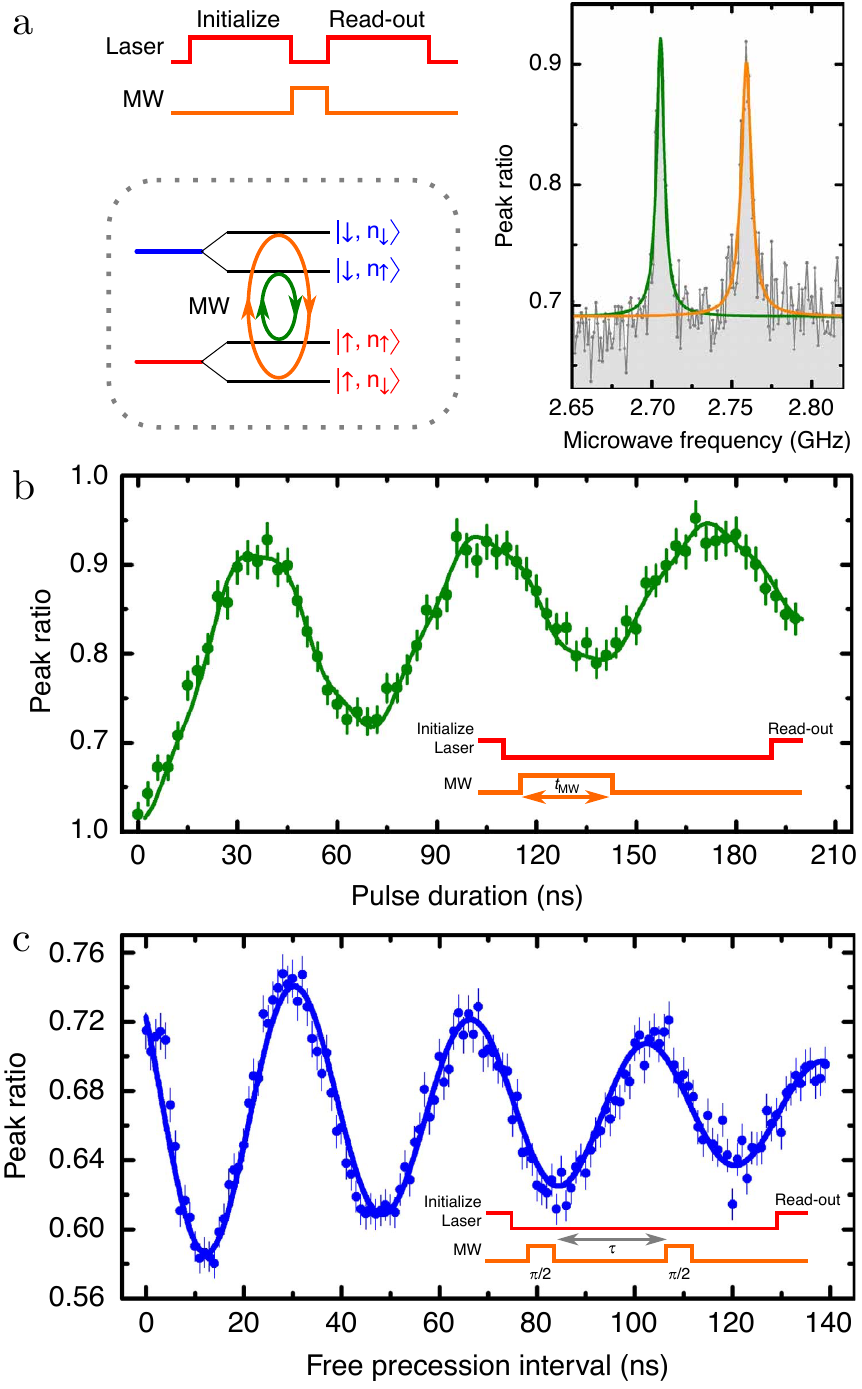}
	\caption{(a) Optically detected magnetic resonance spectrum showing two transitions (green, orange) between the electron spin-flipping and nuclear spin-preserving hyperfine levels of a \siv containing $^{29}$Si \cite{pingault2017}. (b) Microwave-driven Rabi oscillations of the \siv ground state electron spin, featuring a Rabi frequency of about 15\,MHz \cite{pingault2017}. (c) Microwave-driven Ramsey interference between two ground state spin sublevels of the SiV\textsuperscript{-} \cite{pingault2017}.}
	\label{fig5}
\end{figure}
In \cite{pingault2017} the authors implement this by first showing optically detected magnetic resonance (ODMR). To do so, the spin is first optically pumped into one of the spin levels using a resonant laser pulse. Then a microwave pulse is applied which, if on resonance with a magnetic transition, transfers back population into the initial state. This population can then be read out optically by applying a second resonant laser pulse. This sequence is shown in Fig.\,\ref{fig5}(a). In this particular experiment, the authors used an \siv center containing $^{29}$Si carrying a nuclear spin of I=1/2, leading to a hyperfine splitting of the electron spin sublevels. Hence, two distinct resonances are visible in the ODMR spectrum, corresponding to the two electron spin flipping and nuclear spin preserving transitions (orange and green), split by the hyperfine constant A$_{||}$=70\,MHz \cite{rogers2014,edmonds2008,goss2007,gali2013}. Hence this technique also gives access to the potentially long-lived nuclear spin degree of freedom which is not accessible with resonant optical techniques since the hyperfine splitting is much smaller than the lifetime limit of the optical transitions. By applying microwave pulses of variable length and fixed amplitude, the spin can then be coherently driven leading to Rabi oscillations visible in the fluorescence of the readout pulse as shown in Fig\,\ref{fig5}(b). This demonstrates coherent rotations around the x  (or y) axis on the Bloch sphere. To achieve full control a rotation around the z axis is required to set an arbitrary phase of the superposition. This can be realized by harnessing the Larmor precession of the spin in the equatorial plane of the Bloch sphere in a Ramsey interference experiment. To achieve this, first an equal superposition of both spin states is created by applying a first $\pi$/2 microwave pulse, bringing the state in the equatorial plane of the Bloch sphere where it is then allowed to freely evolved for a defined time $\tau$. After this time, a second microwave pulse can then be used to project the rotated state onto the two basis states, resulting in a fluorescence signal depending on the acquired phase as shown in Fig.\,\ref{fig5}(c). In this particular experiment the authors used a simple loop antenna in a distance of about 20\,\textmu s to the \siv center, resulting in a Rabi frequency of about 15\,MHz. Hence, a microwave pulse of about 40\,ns length is required for a $\pi$ pulse, i.e. to completely flip the state of the system from one spin level to the other. The spin coherence time for the emitter investigated there, as extracted from the Ramsey interference measurements, amounts to T$_2^*$=115(9)\,ns, nicely matching the T$_1$-limit of 2T$_1^\text{orbit}$=133(4)\,ns and confirming the phonon-driven decoherence mechanism discussed above. The length of the microwave $\pi$ pulses in this experiment is only about three times shorter than the spin coherence of this \siv and thus significantly limits the application of more complex pulse sequences containing several pulses such as spin echo or dynamic decoupling techniques \cite{suter2016}. While this value can be slightly improved using more elaborate antenna designs, the speed of such microwave-based control schemes is typically limited to the ns range, a significant disadvantage especially in solid state systems where control has oftentimes to be achieved in the presence of fast decoherence processes due to interactions of the system with its host lattice. A particular disadvantage arising from the \siv electronic structure is the fact that each spin sublevel, due to the spin-orbit coupling, also contains an orbital component. While the flip of the electron spin is magnetically allowed, the flip of the orbital component is not, leading to relatively small transition dipole moments compared to e.g. magnetic transitions in the NV. This further limits Rabi frequencies and requires high microwave powers, potentially inducing heating in low temperature experiments. Additionally, the limited spatial selectivity of these techniques can lead to unwanted cross-talk in multi-qubit systems.
\subsection{All-optical control}
Both of the above mentioned limitations can be overcome by using optical fields to realize the coherent control. Optical fields can be selectively focused onto individual qubits with spatial resolutions of only a few hundred nm, requiring only small qubit spacings to avoid cross-addressing several qubits. Moreover, individual qubits can potentially be addressed via photonic structures such as waveguides. Moreover, in systems with e.g. large splittings between the individual electronic levels, ultrashort laser pulses can be used to achieve control on the pico- or femtosecond time scale. Even sub-cycle control, i.e. control on time scales shorter than the qubit frequency splitting, can be achieved using pulses spanning several optical transitions in a $\Lambda$- or $V$-type (one ground state linked to two excited states) level configuration \cite{fuchs2009}. Due to its very large spin orbit splittings between the ground and excited states and the four optical transitions linking these states, the \siv is an ideal candidate to implement this type of control.
\paragraph{Ultrafast resonant control}
Such ultrafast state manipulation has been demonstrated by realizing full coherent control over the center's orbital degree of freedom at zero magnetic field using pulses of only a few ps length \cite{becker2016}.
\begin{figure*}[htb]%
	\includegraphics*[width=\textwidth]{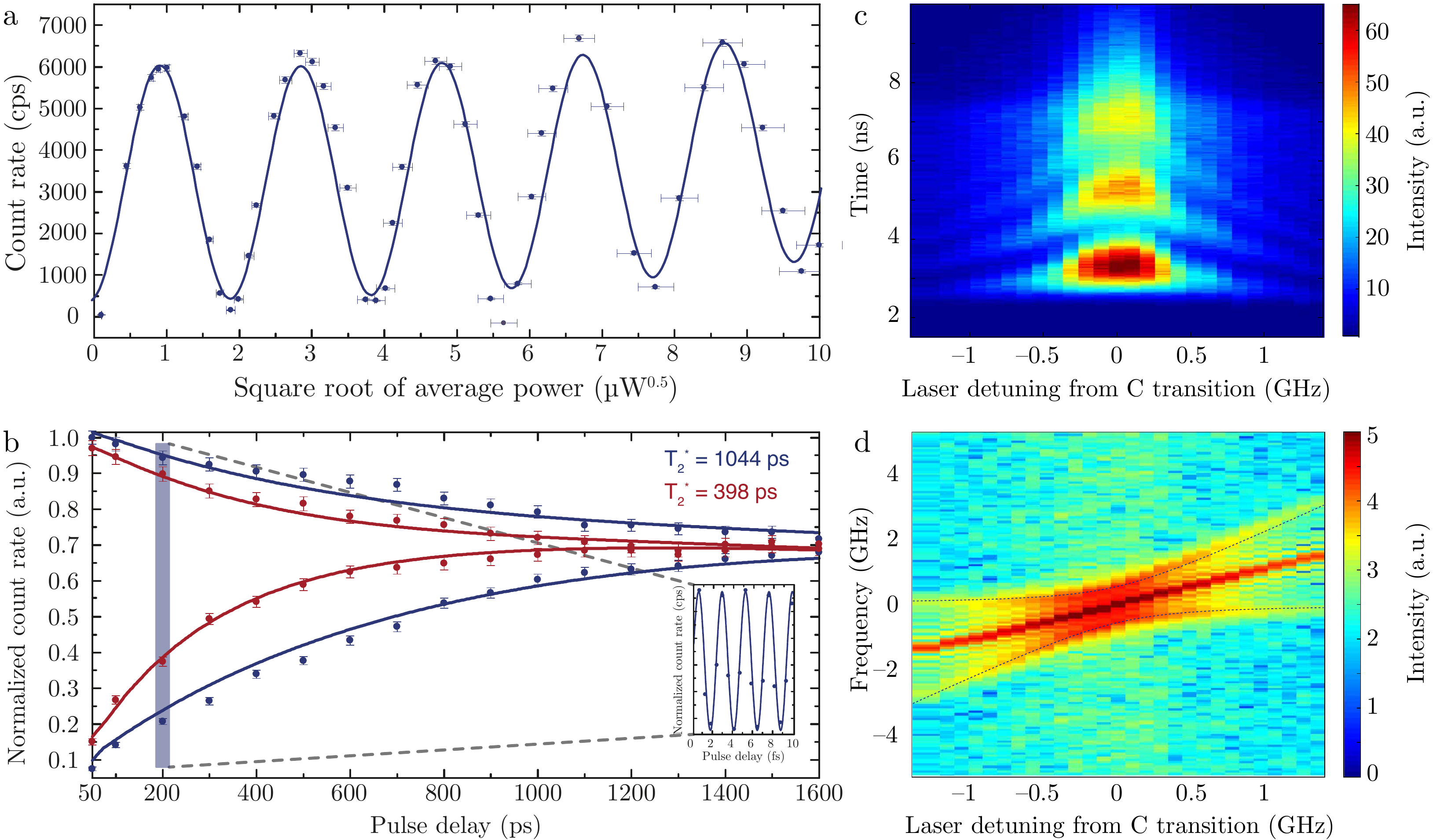}
	\caption{Resonant optical coherent control experiments: (a) and (b): single \siv driven by ps laser pulses; (c) and (d): single \siv driven by modulated cw laser. (a) Resonant optical Rabi oscillations driven by applying 12\,ps long ultrafast laser pulses to transition C \cite{becker2016}. (b) Resonant optical Ramsey interference using two 12\,ps long $\pi$/2 pulses with variable delay applied to transition C (blue) and transition B (red) \cite{becker2016}. (c) Optical Rabi oscillations using a modulated cw laser and fast detection in the time domain for a number of laser detunings \cite{zhou2017}. (d) Fourier transform of (c) indicating a Mollow triplet in frequency space \cite{zhou2017}. 
	}
	\label{fig6}
\end{figure*}
In a first scheme the authors demonstrated full coherent control over a qubit consisting of an orbital ground and excited state level by resonantly addressing individual optical transitions using 12\,ps long pulses. Here, Rabi oscillations are driven by applying pulses with a fixed ultrashort length and variable amplitude. Large rotation angles of up to 10$\pi$ with high contrast and no loss of visibility have been achieved (see Fig.\,\ref{fig6}(a)). More importantly, the authors observe no sign of ionization of the \siv by the high optical peak powers of the ultrashort pulses as it e.g. occurs for NV centers \cite{bassett2014}. Additionally the authors also realize axis control by applying two subsequent ultrashort $\pi$/2 pulses with variable pulse delay resulting in interference patterns with envelopes shown in Fig.\,\ref{fig6}(b) for transition C (blue) and B (red) as well as ultrafast interference fringes on the optical time scale shown in the insert of the figure. From the decay of these patterns the excited state orbital coherence times can be calculated resulting in T$_2^* (\ket{3})$=1044\,ps for the lower and T$_2^*(\ket{4})$=398\,ps for the upper excited state, both limited by the spontaneous decays into the ground states and, in case of $\ket{4}$, by a fast decay into $\ket{3}$. An alternative method to realize optical Rabi oscillations using regular ns long pulses from a modulated cw laser has been presented in \cite{zhou2017}. There, fast superconducting detectors have been used to directly detect the oscillations in the time domain and for different detuning from resonance as depicted in Fig.\,\ref{fig6}(c). The Fourier transform of this measurement is shown in Fig.\,\ref{fig6}(d) indicating a so-called Mollow triplet in which the transition is split by an AC Stark shift (also known as Autler-Townes splitting) in the presence of strong optical fields.
\paragraph{Transition dipole moment \& quantum efficiency}
We here further expand the analysis presented in \cite{becker2016} by utilizing the Rabi oscillation shown in Fig.\,\ref{fig6}(a) to calculate the quantum efficiency $\Phi$ and the transition dipole moment $\mu$ for the \siv center in bulk diamond used there \cite{beckerdiss}.
\begin{figure}[t]%
	\includegraphics*[width=\linewidth]{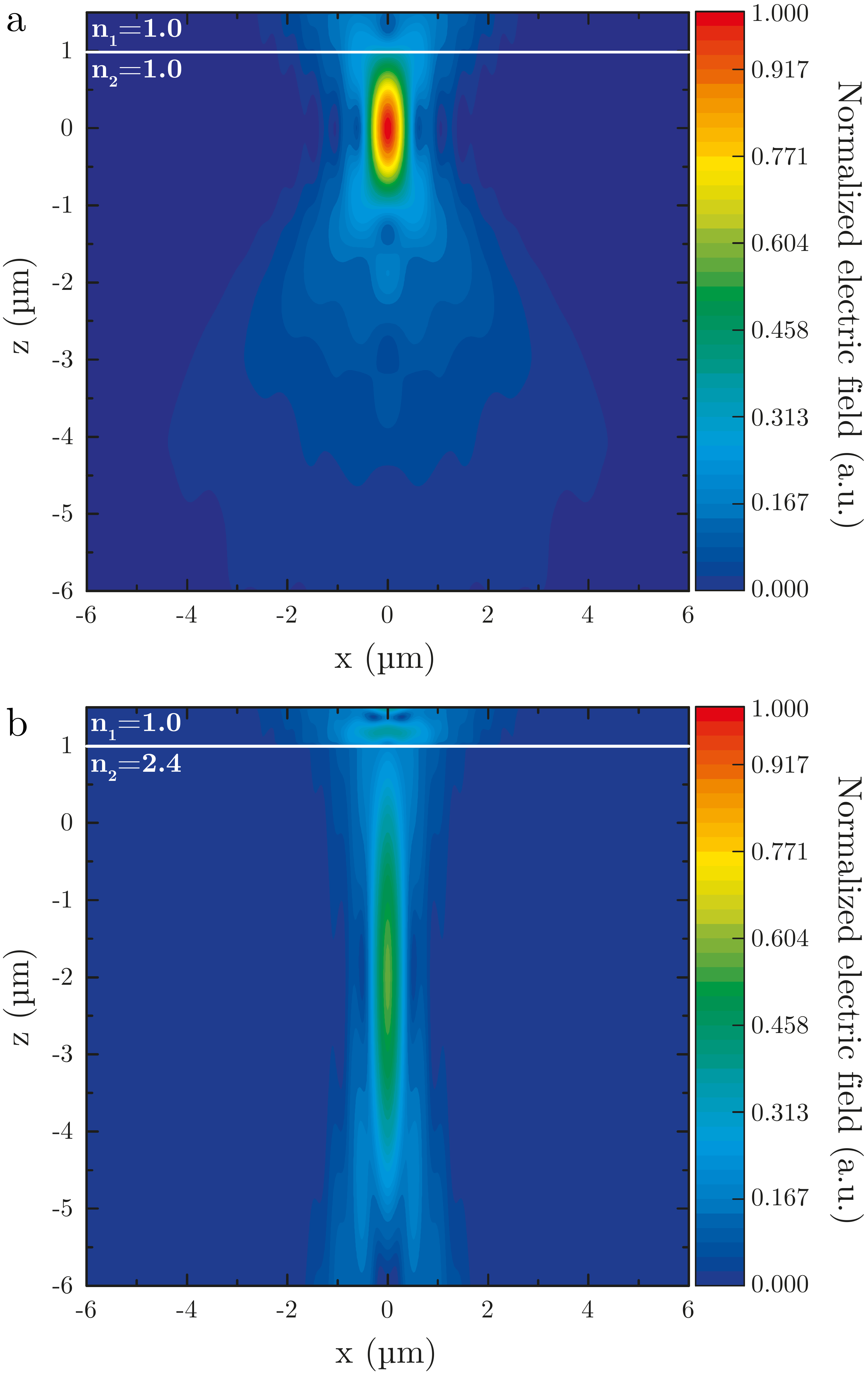}
	\caption{Finite difference time domain (FDTD) simulations of the electric field in the focus of a microscope objective featuring N.A. 0.9 in (a) air and (b) 500\,nm below an air/diamond interface using an input beam diameter of 3\,mm \cite{beckerdiss}.}
	\label{fig7}
\end{figure}
To do so, we calculate the electric field at the coordinates of the emitter: We start with a beam with a Gaussian spatial intensity distribution
\begin{equation}
I(r)=P_0\frac{1}{2\pi\sigma_r^2}e^{-\frac{r^2}{2\sigma_r^2}}
\end{equation}
with $\sigma_r=\frac{d_{focus}}{2\sqrt{2ln(2)}}$, focal diameter $d_{focus}$ and average beam power $P_0$. By assuming that, after careful alignment of the setup, the emitter is centred in the focal spot this simplifies to
\begin{equation}
I(r=0)=\frac{P_0}{2\pi\sigma_r^2}
\end{equation}
and with this expression we can write
\begin{eqnarray}
\label{eq:eint}
\int E(t)dt&=&\sqrt{\frac{P_0\tau TS^2}{\pi\epsilon_0 cn\sigma_r^2} \int_{-\infty}^{\infty}e^{-\frac{2ln(2)}{w_{pulse}}|t|}}\\&=&
\sqrt{\frac{P_0\tau TS^2}{\pi\epsilon_0 cn\sigma_r^2}2 \int_{0}^{\infty}e^{-\frac{2ln(2)}{w_{pulse}}t}}\\&=&
\sqrt{\frac{P_0\tau TS^2}{\pi\epsilon_0 cn\sigma_r^2}\frac{w_{pulse}}{ln(2)}}\nonumber
\end{eqnarray}
for the time-integrated electric field of a single excitation pulse with repetition rate $\tau$ and pulse width $w_{pulse}$, assuming a two-sided exponential for the temporal intensity distribution after filtering by the monolithic Fabry-P\'{e}rot etalons for pulse length adjustment \cite{wolpertdiss}. In this expression we additionally introduce a measured correction factor $T$=0.68 that includes the limited transmission of the microscope objective and cryostat window as well as the back reflection of light at the diamond surface. This value has been measured by determining the back reflected laser power behind the confocal beam splitter. Moreover, as the equations above assume a spherical focal spot we introduce an additional correction factor S, correcting for the increased elongation of the focal spot in diamond compared to air. Since a direct measurement of the geometry of the focal spot is not easily possible, this elongation as well as the focal spot diameter have been extracted from finite difference time domain (FDTD) simulations. In Fig.\,\ref{fig7}(a) the normalized electric field distribution for a Gaussian beam with a diameter of 3\,mm (corresponding to the beam diameter in the setup) focused by a N.A. 0.9 objective lens in air is shown. The resulting focal spot shows only a small elongation and a focal spot diameter of about $d_{focus}^{air}$=893(2)\,nm (FWHM). In contrast to this, Fig.\,\ref{fig7}(b) shows the same beam focused through a diamond/air interface. In this case, we obtain a lateral focus diameter of about $d_{focus}^{diam}$=862(3)\,nm (FWHM). Additionally, a significant axial elongation occurs, caused by spherical aberration due to refractive index mismatch \cite{booth2001}. This results in a noticeable drop in maximum electric field strength in the centre of the focal spot compared to the focus in air.
To take this axial distortion into account we therefore use the ratio S=0.57 of the electric field values in the centres of the focal spots in diamond and air as an additional correction factor. From the Rabi curve in Fig.\,\ref{fig6}(a) we extract a $\pi$ pulse power of P$_{\pi}$=817(16)\,nW and with Eq.\,\ref{eq:eint} we can then calculate the respective transition dipole moment:
\begin{equation}
\mu_C=\frac{\pi\hbar}{\int E(t)dt}=14.3(2)\,\text{Debye}
\end{equation}
Using the definition of the Einstein coefficient A$_{21}$ for spontaneous emission \cite{hilborn1982}
\begin{equation}
A_{21}=\frac{8\pi^2\nu^3}{3\epsilon_0\hbar c^3}\cdot\mu^2
\end{equation}
we can then calculate the natural radiative lifetime $\tau_0$ of the system which amounts to
\begin{equation}
\tau_0=\frac{1}{A_{21}}=6.24\,\text{ns}.
\end{equation}
By comparing this lifetime to the fluorescence lifetime of $\tau=1.85$\,ns measured in \cite{becker2016} via TCSPC, we can then obtain the quantum efficiency of the system which amounts to
\begin{equation}
\Phi=\frac{\tau}{\tau_0}=\frac{1.85\,ns}{6.24\,ns}=29.6(7)\%.
\end{equation}
This value is in good agreement with theoretical and experimental quantum efficiencies previously reported in \cite{riedrich2014}. In there, quantum efficiencies between 15\% and 67\% have been reported for \siv centres in bulk diamond. Moreover, quantum efficiencies below 10\% have been reported for \siv centres in nanodiamonds \cite{neu2012}.
\begin{figure*}[htb]%
	\includegraphics*[width=\textwidth]{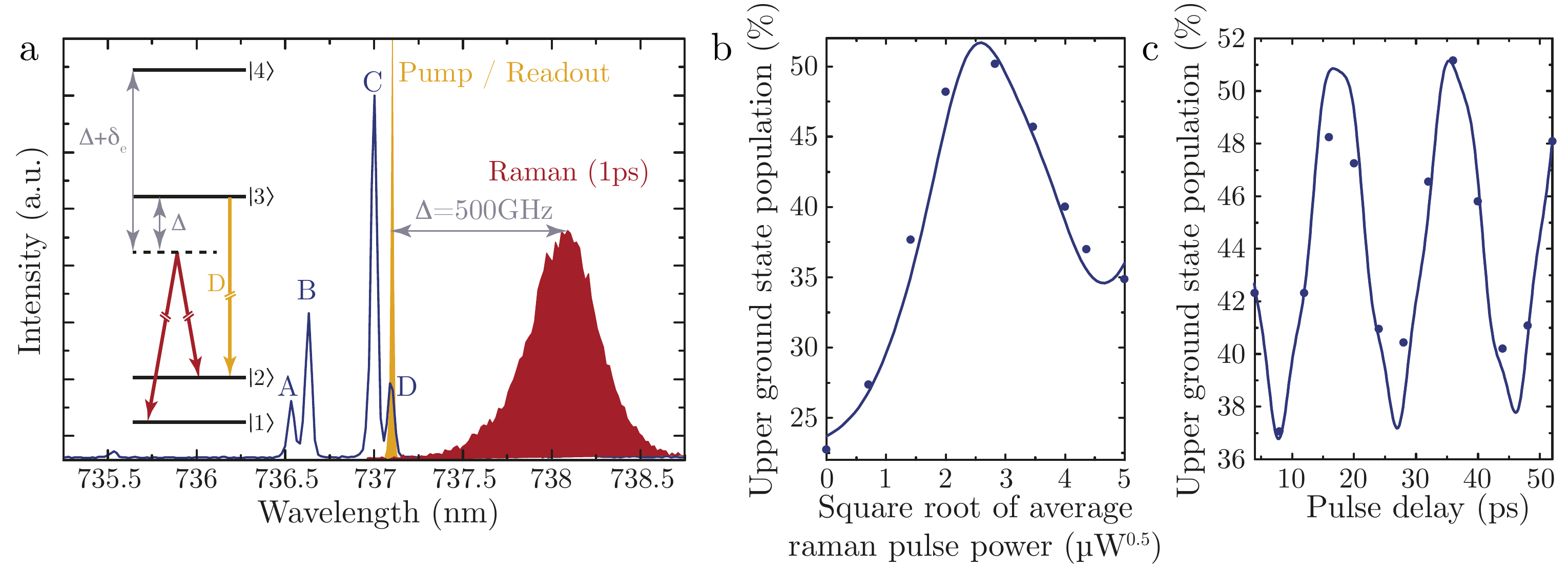}
	\caption{(a) Spectral arrangement of off-resonant Raman laser (red) and resonant pump/readout laser (yellow) relative to the \siv ZPL spectrum \cite{becker2016}. (b) Two-photon Rabi oscillation between orbital ground state levels of the \siv driven by a single off-resonant Raman pulse in a sub-cycle scheme. Rotation angles are limited by available laser power \cite{becker2016}. (c) Two-photon Ramsey interference between orbital ground state levels of the SiV\textsuperscript{-}\cite{becker2016}.
	}
	\label{fig8}
\end{figure*}
Taking the reduced photonic density of states in nanodiamonds into account, these values are consistent with the values reported here. We would like to note that the value of $\Phi$ determined here can be considered a lower bound as the errors above are solely calculated from the power fluctuations of the laser and the calculation assumes an otherwise perfect focus and an ideal placement of the emitter inside the focal spot. In reality, a slightly defocussed beam or a deformation of the focal spot due to surface geometry and roughness might lead to an additional reduction of the electric field strength at the position of the emitter and thus to a higher actual quantum efficiency than the one estimated here.
\paragraph{Raman-based ground state control}
While the resonant optical control is robust and easy to implement as it only requires a single optical field, its value is limited since the coherence time of superpositions consisting of ground and excited state levels is inherently limited by the excited state lifetime. Hence, a qubit solely based on ground state levels is desirable. Optical control of such a qubit consisting of the two orbital ground state levels of the \siv has also been demonstrated in \cite{becker2016} by utilizing two-photon Raman transitions, coupling both ground states in a strongly detuned $\Lambda$-configuration to avoid population of the excited state. In this particular case the authors even implement a sub-cycle control using a single 1\,ps long pulse with a bandwidth of about 300\,GHz simultaneously coupling both arms of the $\Lambda$-scheme with a detuning of $\Delta$=500\,GHz from the lower excited state. An additional modulated cw laser resonant with transition D is used to initialize the system in the lower ground state and to read out the population in the upper ground state after the Raman transfer. The arrangement of the lasers relative to the \siv transitions is depicted in Fig.\,\ref{fig8}(a). With this experimental configuration the authors again show angular control by demonstrating a two-photon Rabi oscillation, displayed in Fig.\,\ref{fig8}(b). In this experiment the rotation angle is mainly limited to 2$\pi$ by the available laser power, since the large detuning implies small two-photon Rabi frequencies $\Omega=(\Omega_C\Omega_D)/2\Delta$ with one-photon Rabi frequencies $\Omega_{C,D}$. Additional limitations to the observed visibilities arise from unwanted resonant excitation with the high-energy tail of the Raman pulse, a problem that can be remedied in future control schemes by moving from sub-cycle control to a Raman control scheme featuring to individual fields with well-controlled bandwidths for both optical transitions. Analogously, by applying subsequent pulses, the authors finally realize full optical control over the \siv ground state by demonstrating the Ramsey interference shown in Fig.\,\ref{fig8}(c) with an oscillation frequency corresponding to the ground state splitting of $\delta_g$=2$\pi\cdot$48\,GHz.\\
While all optical control experiments presented for the \siv so far focussed on controlling its orbital degree of freedom, the very same techniques can be applied in future experiments to also realize full control over its spin degree of freedom. While similar control schemes in the NV center require the application of well-defined external perturbations such as electric field or strain \cite{bassett2014} to induce a spin mixing in the excited state which allows to optically address transitions between spin sublevels of opposite spin projection, such fields are not required in the SiV\textsuperscript{-}, greatly simplifying optical spin control. This is enabled by the above mentioned difference in quantization axes of ground and excited state spin sublevels which occurs for arbitrary magnetic fields misaligned with the center's high symmetry axis. Moreover, by applying strong fields large Zeeman splittings of several tens of GHz can be achieved, enabling ultrafast optical spin control. This is feasible since the coherence times, while being shortened for intermediate field strengths due to spin mixing introduced by the spin orbit interaction, fully recover in the high-field limit as shown in \cite{pingault2014}. Moreover, the Raman transitions can potentially also be used to optically access the nuclear spin degree of freedom since the linewidth of the two-photon transition is defined by the ground state coherence and thus is much smaller than the hyperfine splitting discussed above.

\section{Conclusions \& Future perspectives}
In this article we provided a compact overview over the electronic structure and coherence properties of the \siv center in diamond and reviewed the only recently demonstrated experiments implementing microwave-based as well as all-optical coherent control of the \siv center's orbital and spin degree of freedom. These techniques constitute a powerful toolbox for a number of exciting future applications such as efficient spin-photon interfaces \cite{togan2010}. Moreover, while the above discussed experiments demonstrate the accessibility of a full set of single qubit gates, for QIP applications a universal two-qubit gate such as the CNOT needs to be realized in the future. A promising step towards this has recently been made by demonstrating the entanglement of two individual \siv centers embedded in a common nanophotonic waveguide via indistinguishable Raman photons emitted from both centers \cite{sipahigil2016}. The combination of narrow bandwidth and small Huang-Rhys factor of the \siv allows efficient coupling to such photonic structures and hence raises hope for high entanglement rates in probabilistic entanglement schemes \cite{cabrillo1999}, a key requirement for a scalable quantum information processing technologies.\\
Further exciting future experimental directions arise from the narrow inhomogeneous broadenings that can be achieved even in dense \siv ensembles. This feature is unique for solid-state systems which oftentimes suffer from severe broadening due to inhomogeneities in their host material and is enabled by the inversion symmetry of the defect, rendering it insensitive against first-order Stark and strain shifts. By utilizing the above discussed Raman transitions in such dense ensembles the realization of efficient and potentially integratable quantum memories \cite{kozhekin2000}, capable of storing single photons with bandwidths of several tens of GHz appear feasible. Moreover, these ensembles open the possibility to explore a range of applications relying on single photon nonlinearities, such as single photon intensity \cite{harris1998} or phase switching \cite{liu2016}.\\
Ultimately it is clear that, while the \siv offers extraordinary spectral properties, future experiments will also have to focus on prolonging its electron spin coherence times or on addressing the potentially long-lived nuclear spin to extend its applicability in quantum information technologies, rendering its challenges diametrically opposed to the ones of the NV center. In any case, the experiments reviewed here significantly deepen our understanding of the physics of color centers in diamond and, for the first time, now also enable a targeted search for new color centers with optimized properties for QIP applications combining the spin characteristics of the NV with the optical features of the SiV\textsuperscript{-}.

\begin{acknowledgement}
We thank Laura Kreiner and Philipp Fuchs for providing the FDTD simulations presented in Fig\,\ref{fig4} and Fig.\,\ref{fig6}, respectively.
\end{acknowledgement}

%
\providecommand{\WileyBibTextsc}{}
\let\textsc\WileyBibTextsc
\providecommand{\othercit}{}
\providecommand{\jr}[1]{#1}
\providecommand{\etal}{~et~al.}

%

\end{document}